\RequirePackage{ifpdf}
\ifpdf 
\documentclass[pdftex]{sigma}
\else
\documentclass{sigma}
\fi

\begin{document}

\allowdisplaybreaks

\renewcommand{\thefootnote}{$\star$}

\renewcommand{\PaperNumber}{075}

\FirstPageHeading

\ShortArticleName{Supersymmetrical Separation of Variables in Two-Dimensional Quantum Mechanics}

\ArticleName{Supersymmetrical Separation of Variables\\ in Two-Dimensional Quantum Mechanics\footnote{This
paper is a contribution to the Proceedings of the Workshop ``Supersymmetric Quantum Mechanics and Spectral Design'' (July 18--30, 2010, Benasque, Spain). The full collection
is available at
\href{http://www.emis.de/journals/SIGMA/SUSYQM2010.html}{http://www.emis.de/journals/SIGMA/SUSYQM2010.html}}}

\Author{Mikhail V.~IOFFE}


\AuthorNameForHeading{M.V.~Iof\/fe}

\Address{Saint-Petersburg State University, St.-Petersburg, 198504 Russia}
\Email{\href{mailto:m.ioffe@pobox.spbu.ru}{m.ioffe@pobox.spbu.ru}}
\URLaddress{\url{http://hep.niif.spbu.ru/staff/ioffe_e.htm}}

\ArticleDates{Received August 24, 2010, in f\/inal form September 19, 2010;  Published online September 24, 2010}

\Abstract{Two dif\/ferent approaches are formulated to analyze two-dimensional quantum models
which are not amenable to standard separation of variables.
Both methods are essentially based on supersymmetrical second order intertwining relations and
shape invariance~-- two main ingredients of the supersymmetrical quantum mechanics.
The f\/irst method explores the opportunity to separate variables in the supercharge,
and it allows to f\/ind a part of spectrum of the Schr\"odinger Hamiltonian.
The second method works when the standard separation of variables procedure
can be applied for one of the partner Hamiltonians. Then the spectrum and wave functions
of the second partner can be found. Both methods are illustrated by the example of two-dimensional
generalization of Morse potential for dif\/ferent values of parameters.}


\Classification{81Q60}

\section{Introduction}

The exactly solvable models in quantum mechanics are of special interest during many years both by
methodological and practical reasons. By now, the main achievements were related to one-dimensional Shr\"odinger
equation. Indeed, a list of exactly solvable one-dimensional problems (Harmonic oscillator, Coulomb, Morse,
P\"oschl--Teller potentials etc.) was obtained by an algebraic procedure in the framework of factorization
method \cite{infeld} in the middle of last century. This method was reproduced rather recently in supersymmetrical
quantum mechanics approach~\cite{cooper}
initiated by the seminal papers of E.~Witten~\cite{witten}. More of that, this approach gave many new exactly
solvable potentials which were obtained as superpartners of ``old'' exactly solvable models. It is necessary to
mention also the important paper of L.~Gendenstein~\cite{gendenstein}, where the
new fruitful notion of shape invariance was introduced. For the sake of truth, more than a century ago the so
called Darboux transformation~\cite{darboux} for Sturm--Liouville equation was well known among mathematicians.
Its application to a specif\/ic Schr\"odinger-like equation is actually equivalent \cite{abi,bagrov}
to the factorization method.

The situation is much worse for two-dimensional quantum mechanics. The only regular method to solve analytically
the Schr\"odinger equation is well known method of separation of variables~\cite{miller}. This method replaces
the two-dimensional problem by a pair of one-dimensional problems. It can be used for very restrictive class of models.
Full classif\/ication of models which allowed separation of variables was given by L.P.~Eisenhart~\cite{eisenhart}:
four possibilities exist -- Cartesian, polar,
elliptic and parabolic coordinates. The general form of potentials amenable to separation of variables
is known explicitly up to arbitrary functions of one variable. And analytical solution is possible only if these
functions belong to the list of exactly solvable potentials. All these Hamiltonians $H$ are integrable: the
symmetry operator $R$ of
second order in derivatives (in momenta) exists: $[H, R]=0$.
Besides models amenable to separation of variables, the class of so called Calogero-like
models \cite{calogero} is known as well. They describe the specif\/ic forms of pairwise interaction
of $N$ particles on a line, and they are solvable by means of special transformation of variables which
leads to a separation of variables. The problem is in analogous state for higher dimensions of space.

Intermediate class of models~-- quasi-exactly-solvable (QES) potentials (or, partially sol\-vable)~-- became
interesting during last years. This notion concerns models for which only a part of spectrum and corresponding
wave functions can be found analytically.
In one-dimensional quantum mechanics a lot of such models were built with some hidden algebraic
structure~\cite{turbiner}. The supersymmetrical approach also gave some new QES potentials~\cite{raj}.

Thus, the search of new approaches to solution of nontrivial two-dimensional quantum mechanical models
seems to be of current importance. It was already mentioned above that supersymmetrical quantum mechanics
provides both new ways to derive some old results and interesting method to obtain new ones. In this paper
we shall focus on the case of two-dimensional Schr\"odinger equation. Namely, we shall present two procedures
of using of the supersymmetrical intertwining relations with supercharges of second order in derivatives
as procedures of SUSY-separation of variables.

In Section~\ref{section2} the general form of two-dimensional supersymmetrical quantum mechanics with second order
supercharges will be formulated.
Section~\ref{section3} presents the f\/irst procedure of SUSY separation of variables where variables are separated
in the supercharge. It leads to QES models, and the specif\/ic model of two-dimensional Morse potential
illustrates this method. In Section~\ref{section4} the second procedure of SUSY separation of variables is given
where variables are separated in one of partner Hamiltonians. In the case of the same Morse model,
but with particular values of parameter, it allows to solve the model completely, i.e.\ to f\/ind analytically
the whole spectrum and all wave functions.

\section{Two-dimensional SUSY quantum mechanics}\label{section2}

Direct generalization of one-dimensional Witten's SUSY quantum mechanics to the arbitrary dimensionality $d$ of
space was
formulated in \cite{abi, abei}. The Superhamiltonian included $(d+1)$ matrix components of
dif\/ferent matrix dimensionality, and these components are intertwined by components of supercharge -- operators
linear in derivatives.
In particular, in the case of $d=2$ two scalar Hamiltonians and a $2\times 2$ matrix Hamiltonian are intertwined
\begin{gather*}
H^{(0)} \Longleftrightarrow H_{ik} \Longleftrightarrow \widetilde H^{(0)} \nonumber\\
\quad\quad\,\,\, q_i^{\pm} \quad\quad\,\,\,\,\, p_i^{\mp}\nonumber
\end{gather*}
where
\[
q_i^{\pm}=\mp\partial_i+(\partial_iW(\vec x)), \qquad  p_i^{\pm}=\varepsilon_{ik}q_k^{\mp},\qquad
  H^{(0)}q_i^+=q_k^+H^{(1)}_{ki},\qquad
  \widetilde H^{(0)}p_i^+=p_k^+H^{(1)}_{ki}.
\]

Some physical problems were considered in this framework. For example, the spectrum of the Pauli operator
describing spin $1/2$ fermion in the external electrostatic and magnetic f\/ield was
investigated \cite{pauli, neelov,iknn}. However, the following natural question arises:
is it possible to avoid matrix Hamiltonians
from the scheme? Any attempt to intertwine two scalar Hamiltonians by means of f\/irst order operators leads to
potentials with standard separation of variables~\cite{kuru} which are not interesting for us here.

The nontrivial way to avoid matrix Hamiltonians lies in the framework of polynomial SUSY. The latter was used
for the f\/irst time \cite{spiridonov,bagrov} in one-dimensional SUSY quantum mechanics. In two-dimensional context it was proposed in \cite{david},
where a pair of scalar two-dimensional
Hamiltonians $H^{(0)}$, $H^{(1)}$ was intertwined by second order operators $Q^{\pm}$
\[
H^{(0)}Q^+=Q^+H^{(1)},\qquad Q^-H^{(0)}=H^{(1)}Q^-,
\]
where the Hamiltonians have the Schr\"odinger form
\[
H^{(0,1)}=-\partial^2_i+V^{(0,1)}(\vec x).
\]
As for the intertwining supercharges $Q^{\pm}$,
the f\/irst naive idea is to choose reducible (factorized) supercharge $Q^+=q_i^+\widetilde q_i^- $
\begin{gather*}
\quad\,\quad\,\,\, q_i^{\pm} \quad\quad\quad p_i^{\mp}\nonumber\\
H^{(0)} \Longleftrightarrow H_{ik} \Longleftrightarrow \widetilde H^{(0)} \nonumber\\
\,\,\quad\quad\,\quad\quad\,\parallel \nonumber\\
H^{(1)} \Longleftrightarrow H_{ik} \Longleftrightarrow \widetilde H^{(1)} \nonumber\\
\,\,\,\quad\quad\, \widetilde q_i^{\pm} \quad\quad\quad \widetilde p_i^{\mp}\nonumber
\end{gather*}
It is too naive \cite{david}, since this construction leads to Hamiltonians amenable to separation of variables
in polar coordinates
\[
V(\vec x)= a^2\rho^2+\frac{1}{\rho^2}F(\varphi).
\]

The second idea is to choose $Q^+=q_i^+U_{ik}\widetilde q_k^-$ with some unitary twist by constant
matrix $U_{ik}$
\begin{gather*}
\quad\quad\,\,\,\, q_i^{\pm} \quad\quad\quad\,\,\,\quad p_i^{\mp}\nonumber\\
H^{(0)} \Longleftrightarrow \,\,\, H_{ik} \,\quad\Longleftrightarrow \widetilde H^{(0)} \nonumber\\
\quad\quad\,\,\quad\quad\quad\parallel\nonumber\\
H^{(1)} \Longleftrightarrow U_{il}H_{lm}U^{\dagger}_{mk}
\Longleftrightarrow \widetilde H^{(1)} \nonumber\\
\quad\quad\,\,\,\, \widetilde q_i^{\pm} \quad\quad\quad\quad\quad\,\quad \widetilde p_i^{\mp}\nonumber
\end{gather*}
Some QES models were obtained by this trick \cite{valinevich}.

The most general form of second order supercharges
\[
Q^+= g_{ik}{(\vec x)}\partial_i \partial_k + C_i(\vec x )\partial_i +
B(\vec x),\qquad  Q^-\equiv (Q^+)^{\dagger}
\]
leads to a complicate system of nonlinear second order dif\/ferential equations
for functions $g_{ik}$, $C_i$, $B$, and potentials $V^{(0,1)}(\vec x)$.
Its general solution is impossible, but some particular solutions were found \cite{david, innn}.
The simplest choice $g_{ik}(\vec x)=\delta_{ik}$ gives the separation of variables in polar coordinates.
The Lorentz form $g_{ik}(\vec x)={\rm diag}(1,-1)$ does not lead to separation of variables, and some
particular solutions were found~\cite{david}.

Here we focus on $g_{ik}(\vec x)={\rm diag}(1,-1)$. In this case, the system is simplif\/ied essentially.
New variables $x_{\pm}=x_1\pm x_2$ are useful together with $x_1$, $x_2$.
Using the intertwining relations, one can prove that new functions $C_{\pm}$ depend on one variable only
\[
C_+ \equiv C_1 - C_2=C_+(x_+),
\qquad  C_- \equiv C_1 + C_2=C_-(x_-), \qquad  x_{\pm}=x_1 \pm x_2.
\]
The general solution for Lorentz metric can be provided by solving the only equation
\[
\partial_-(C_- F) = -\partial_+(C_+ F),
\]
where new useful function is $ F=F_{1}(x_{+}+x_{-}) + F_{2}(x_{+}-x_{-})$.
Thus, the equation is the functional dif\/ferential equation, and no regular procedure of its solution is known.

The required potentials $ V^{(0,1)}(\vec x)$ and the function $ B(\vec x) $ are expressed in terms
of $C_{\pm}$ and~$F_{1,2}$
\begin{gather*}
V^{(0,1)}=\pm\frac{1}{2}(C_+' + C_-')+\frac{1}{8}\big(C_+^2 + C_-^2\big) + \frac{1}{4}\left( F_2(x_+
-x_-) - F_1(x_+ + x_-)\right), \\
B=\frac{1}{4}\left( C_+ C_- + F_1(x_+ +
x_-) + F_2(x_+ - x_-)\right).
\end{gather*}
A variety of such pairs of potentials was found in \cite{david}.

\section{SUSY-separation of variables I: QES models}\label{section3}

The f\/irst variant of SUSY-separation of variables is realized when the Hamiltonian $H$ does not
allow standard separation of variables, but the supercharge $Q^+$
does allow \cite{new,ioffe}.
The general scheme is the following. Let's suppose that we know zero modes of $Q^+$
\[
Q^+\Omega_n (\vec x)=0,\quad n=0,1,\dots,N,\qquad
Q^+ \vec\Omega (\vec x)=0.
\]
The intertwining relation
\[
H^{(0)}Q^+=Q^+H^{(1)}
\]
obey the important property: the space of zero modes is closed under the action of $H^{(1)}$:
\[
H^{(1)}\vec\Omega (\vec x) = \hat C \vec\Omega (\vec x).
\]

If the matrix $\hat C$ is known, and if it can be diagonalized
\[
 \hat B \hat C = \hat\Lambda \hat B,\qquad  \hat\Lambda =
 {\rm diag}(\lambda_0,\lambda_1,\dots,\lambda_N),
\]
the eigenvalues of $H^{(1)}$ can be found algebraically
\[
H^{(1)} (\hat B\vec\Omega (\vec x)) = \hat\Lambda(\hat B\vec\Omega (\vec x)).
\]
Thus, for realization of this scheme we need
\begin{enumerate}
\itemsep=0pt
\item[--] to f\/ind zero modes $\Omega_n(\vec x)$;

\item[--] to f\/ind constant matrix $B$, such that $\hat B \hat C = \hat\Lambda \hat B$.
\end{enumerate}

As for zero modes, they can be obtained by using the special similarity transformation (not unitary!), which
removes the terms linear in derivatives from $Q^+$
\begin{gather*}
q^+ = e^{-\chi (\vec x)} Q^+ e^{+\chi (\vec x)} =
 \partial_1^2 -
\partial_2^2 + \frac{1}{4}(F_1(2x_1) + F_2(2x_2)),
\\
\chi (\vec x) = -\frac{1}{4}\left( \int
C_+(x_+)dx_+ + \int C_-(x_-)dx_- \right).
\end{gather*}
Now, $q^+$ allows separation of variables for arbitrary
solution of intertwining relations, and we obtain the f\/irst variant of new procedure~-- SUSY-separation of variables.
Similarly to the conventional separation of variables, separation of variables in the operator $q^+$ itself
does not guarantee solvability of the problem.

The next task is to solve two one-dimensional problems
\begin{gather*}
\left(-\partial_1^2
-\frac{1}{4}F_1(2x_1)\right)\eta_n(x_1)=\epsilon_n\eta_n(x_1),\\
\left(-\partial_2^2
+\frac{1}{4}F_2(2x_2))\rho_n(x_2\right)=\epsilon_n\rho_n(x_2).
\end{gather*}

Three remarks are appropriate now.

\begin{remark}
The same similarity transformation of $H^{(1)}$ does not lead
to operator amenable to separation of variables.
\end{remark}

\begin{remark}
The normalizability of $\Omega_n$ has to be studied attentively
due to non-unitarity of the similarity transformation.
\end{remark}

\begin{remark}
We have no reasons to expect exact solvability of the model, but quasi-exact-solvability
can be predicted.
\end{remark}

As for the matrix $\hat B$, it must be found by some specif\/ic procedure. Such procedure was
used in example which will be presented below.

In principle, the f\/irst scheme of SUSY-separation of variables can
be used for arbitrary models satisfying intertwining relations by supercharges with Lorentz metrics.
The list of solutions of intertwining relations is already rather long, and it may increase in future.
The main obstacle is analytical solvability of one-dimensional equations, obtained after separation of
variables in the operator~$q^+$.

Below we describe brief\/ly such a model which can be considered as the generalized two-dimensional Morse potential
\begin{gather*}
C_+=4a\alpha,\qquad C_-=4a\alpha \coth \frac{\alpha x_-}{2},\nonumber\\
f_i(x_i) \equiv   \frac{1}{4}
F_i(2x_i)=-A\left(e^{-2\alpha x_i} - 2 e^{-\alpha
x_i}\right),\qquad i=1,2,
\\
V^{(0),(1)}= \alpha^2a(2a \mp 1)\sinh^{-2}\left(\frac{\alpha x_-}{2} \right) +
4a^2\alpha^2+ A \left[e^{-2\alpha
x_1}-2 e^{-\alpha x_1} + e^{-2\alpha x_2}-2 e^{-\alpha
x_2}\right],
\end{gather*}
where $A>0$, $\alpha >0$, $a$ is real.

To explain the name, we present the potential in the form
\[
V(\vec x)= V_{\rm Morse}(x_1)+V_{\rm Morse}(x_2)+ v(x_1,x_2),
\]
where f\/irst two terms are just one-dimensional Morse potentials, and the last term
mixes variables~$x_1$, $x_2$.

The solutions of one-dimensional Schr\"odinger equations are well known~\cite{landau},
and the zero modes can be written \cite{new,ioffe} as
\begin{gather*}
\Omega_n(\vec x) = \left(\frac{\alpha}{\sqrt{A}} \frac{\xi_1\xi_2}{|\xi_2 -\xi_1|}\right)
^{2a}\exp\left(-\frac{\xi_1+\xi_2}{2}\right) (\xi_1\xi_2)^{s_n}
 F(-n, 2s_n +1; \xi_1) F(-n, 2s_n +1; \xi_2),
\\
\xi_i\equiv  \frac{2\sqrt{A}}{\alpha}\exp(-\alpha x_i), \qquad
s_n=\frac{\sqrt{A}}{\alpha}-n-\frac{1}{2} > 0.
\end{gather*}
The conditions of normalizability and of absence of the ``fall to the center'' are
\[
a \in
\left({-}\infty,  -\frac{1}{4}-\frac{1}{4\sqrt{2}}\right), \qquad s_n =
\frac{\sqrt{A}}{\alpha}-n-\frac{1}{2} > -2a >0.
\]

To obtain the matrix $\hat C$ explicitly, one must act by $H^{(1)}$ on $\Omega_n$.
The matrix turns out to be triangular, and therefore, the energy eigenvalues
coincide with its diagonal elements
\[
E_k=c_{kk}=-2\big(2a\alpha^2s_k-\epsilon_k\big).
\]

To f\/ind a variety of wave functions is a more dif\/f\/icult task. For that it is necessary to
f\/ind all elements of $\hat C$ and all elements of matrix $\hat B$.
The recurrent procedure for the case of two-dimensional Morse potential was given in \cite{new,ioffe}.
This variety can be enlarged by means of shape invariance property \cite{shape} of the model
\[
H^{(0)}(\vec x; a)=H^{(1)}(\vec x; \tilde a)+ {\cal R}(a),\qquad \tilde a=a-1/2,\qquad
{\cal R}(a)=\alpha^2(4a-1).
\]
Similarly to one-dimensional shape invariance, each wave function constructed by SUSY-sepa\-ra\-tion of variables
leads to a set of additional wave functions
\begin{gather*}
H^{(0)}(a)\left[ Q^-(a)Q^-\left(a-\frac{1}{2}\right)\cdots Q^-\left(a-\frac{M-1}{2}\right)\Psi\left(a-\frac{M}{2}\right) \right] \\
\qquad {}= \left(E_0\left(a-\frac{M}{2}\right)
+ {\cal R}\left(a-\frac{M-1}{2}\right)+ \cdots + {\cal R}(a)\right)\\
\qquad\quad{}\times
\left[ Q^-(a)Q^-\left(a-\frac{1}{2}\right)\cdots Q^-\left(a-\frac{M-1}{2}\right)\Psi\left(a-\frac{M}{2}\right) \right].
\end{gather*}
Analogous approach works for the two-dimensional generalization of P\"oschl--Teller model~\cite{valinevich}
and for some two-dimensional periodic
potentials~\cite{periodic}.

\section{SUSY-separation of variables II: exact solvability}\label{section4}

Among all known solutions of two-dimensional intertwining relations with second order supercharges
a subclass exists \cite{physrev}, where one of intertwined Hamiltonians is amenable to
standard separation of variables due to specif\/ic choice of parameters of the model.
Its superpartner still does not allow separation of variables.

The scheme will be described below for the same specif\/ic model which is two-dimensional
generalization of Morse potential
\[
V^{(0),(1)}=
\alpha^2a(2a \mp 1)\sinh^{-2}\left(\frac{\alpha x_-}{2} \right) +
4a^2\alpha^2 + A \left[e^{-2\alpha
x_1}-2 e^{-\alpha x_1} + e^{-2\alpha x_2}-2 e^{-\alpha
x_2}\right].
\]
Let's choose $a_0=-1/2$ in order to vanish the mixed term in $V^{(1)}$.
Then $H^{(1)}$ allows the conventional separation of variables.
Moreover, after separation of variables each of obtained one-dimensional problems is exactly solvable.
We met just this one-dimensional problem above in a dif\/ferent context.

The discrete spectrum of this one-dimensional model is
\[
\epsilon_n=-\alpha^2s_n^2,\qquad
s_n\equiv\frac{\sqrt{A}}{\alpha}-n-\frac{1}{2} >0,\qquad  n=0,1,2,\ldots .
\]
Wave functions are expressed in terms of degenerate hypergeometric
functions
\[
\eta_n(x_i) = \exp\left(-\frac{\xi_i}{2}\right) (\xi_i)^{s_n}
F(-n, 2s_n +1; \xi_i),\qquad \xi_i\equiv \frac{2\sqrt{A}}{\alpha}\exp(-\alpha x_i).
\]

Due to separation of variables, the two-dimensional problem with $H^{(1)}(\vec x)$
is exactly solvable. Its energy eigenvalues are
\[
E_{n,m}=E_{m,n}=\epsilon_n+\epsilon_m,
\]
being two-fold degenerate for $n\neq m$.
The corresponding eigenfunctions can be chosen as
symmetric or (for $n\neq m$) antisymmetric combinations
\[
\Psi^{(1)\, S,A}_{E_{n,m}}(\vec x) = \eta_n(x_1)\eta_m(x_2)\pm\eta_m(x_1)\eta_n(x_2).
\]

Our aim here is to solve completely the problem for $H^{(0)}(\vec x)$ with $a_0=-1/2$.
The main tool is again the SUSY intertwining relations, i.e.\ isospectrality of
$H^{(0)}$ and $H^{(1)}$ but up to zero modes and singular properties of $Q^{\pm}$.
In general, we may expect three kinds of levels of $H^{(0)}(\vec x)$:
\begin{enumerate}\itemsep=0pt
\item[(i)]  The levels, which coincide with $E_{nm}$. Their wave functions can be
obtained from $\Psi^{(1)}$ by means of $Q^+$.

\item[(ii)] The levels, which were absent in the spectrum of $H^{(1)}(\vec x)$, if some wave functions
of $H^{(0)}(\vec x)$ are simultaneously
the zero modes of $Q^-$. Then the second intertwining relation
would not give any partner state among bound states of $H^{(1)}(\vec x)$.

\item[(iii)] The levels, which were also absent in the spectrum of $H^{(1)}(\vec x)$,
if some wave functions
of~$H^{(0)}(\vec x)$ become nonnormalizable after action of operator $Q^-$.
\end{enumerate}

We have to analyze these three classes of possible bound states of $H^{(0)}$ one after another.

(i) The f\/irst SUSY intertwining relation gives
the two-fold degenerate wave functions of $H^{(0)}$
with energies $E_{nm}$: $\Psi^{(0)}_{E_{nm}}=Q^+\Psi^{(1)}_{E_{nm}}$.
But $Q^+$ includes singularity on the line $x_1=x_2$, therefore the
normalizability of $\Psi^{(0)}_{E_{n,m}}$ depends crucially on the behavior of
$\Psi^{(1)}_{E_{n,m}}$ on the line $\xi_1=\xi_2$. One can check that only antisymmetric
functions $\Psi^{(1)}$ survive, i.e.\ only symmetric $\Psi^{(0)}$ survive.
This fact can be demonstrated~\cite{physrev} both by direct calculation and by indirect method - by means of
symmetry operator $R^{(0)}$.

The indirect method explores that the symmetry operator $R^{(0)}=Q^-Q^+$ for $a_0=-1/2$
can be written in terms of one-dimensional Morse Hamiltonians $h_1(x_1)$, $h_2(x_2)$
\[
R^{(0)}= (h_1(x_1)-h_2(x_2) )^2 + 2\alpha^2 (h_1(x_1)+h_2(x_2) )+\alpha^4.
\]
Therefore,
\[
R^{(0)}\Psi^{(0) A}_{E_{n,m}}(\vec x)=r_{n,m}\Psi^{A}_{E_{n,m}}(\vec x),\qquad
r_{n,m}=\alpha^4\big[(n-m)^2-1][(s_n+s_m)^2 - 1\big],
\]
and
\[
\|\Psi^{(1) S}_{E_{n,m}}\|^2=
\langle\Psi^{(0) A}_{E_{n,m}}\mid Q^-Q^+\mid \Psi^{(0) A}_{E_{n,m}}\rangle=
r_{n,m}\|\Psi^{(0) A}_{E_{n,m}}\|^2.
\]

For $n=m$, wave functions $\Psi^{(0) S}_{E_{n,n}}$ vanish identically by trivial reasons.
It is clear now that wave functions $ \Psi^{(0) S}_{E_{n,n\pm 1}}$ also vanish.
For all other $n$, $m$, functions
$\Psi^{(0) S}_{E_{n,m}}$ have positive and f\/inite norm, and there is no degeneracy of these levels.

(ii) These possible bound states of $H^{(0)}$ are the
normalizable zero modes of $Q^-$.
The variety of such zero modes is known from \cite{new}: they exist only for positive values of $a$
\[
a \in \left(\frac{1}{4}+\frac{1}{4\sqrt{2}}  ,  +\infty \right),
\]
which does not contain the value $a_0=-1/2$. Thus,
no normalizable bound states of this class exist for $H^{(0)}$.

(iii) We have to study an opportunity that $Q^-$ destroys normalizability of some
eigenfunctions of $H^{(0)}$. It could occur due to singular character
of $Q^-$ at $x_1 = x_2$.
The analysis was performed~\cite{physrev} in suitable coordinates. It shows that
$Q^-$ is not able to transform normalizable wave function
to nonnormalizable. Therefore, the third class of possible wave functions
$H^{(0)}$ does not exist too.

Summing up, the spectrum of $H^{(0)}$ with $a_0=-1/2$
consists only
of the bound states with energies $E_{nm}$ for $|n-m|>1$.
This spectrum is bounded from above
by the condition of positivity of $s_n$, $s_m$: $n,m < \sqrt{A}/\alpha - 1/2$.
The corresponding wave functions are obtained analytically~\cite{physrev}.

The results above can be expanded to the whole hierarchy of Morse potentials
with $a_k=-(k+1)/2$ with $k=0,1,\dots$ by means of shape invariance property.
Let's denote elements of the hierarchy as $H^{(0)}(\vec x; a_k)$, $H^{(1)}(\vec x; a_k)$.
All these Hamiltonians are also exactly solvable due to shape invariance of the model
\[
H^{(0)}(\vec x; a_{k-1})=H^{(1)}(\vec x; a_{k}), \qquad  k=1,2,\ldots .
\]
This means that the following chain (hierarchy) of Hamiltonians can be built
\begin{gather*}
H^{(1)}(\vec x; a_0)\div H^{(0)}(\vec x; a_0)=H^{(1)}(\vec x; a_1)\div H^{(0)}(\vec x; a_1)=
\cdots \div H^{(1)}(\vec x; a_{k-1})\\
\phantom{H^{(1)}(\vec x; a_0)\div H^{(0)}(\vec x; a_0)}{}
=H^{(0)}(\vec x; a_{k})\div H^{(0)}(\vec x; a_{k}),
\end{gather*}
where the sign $\div $ denotes intertwining by $Q^{\pm}(a_i)$.

In the general case, the functions
\[
\Psi^{(0)}_{E_{n,m}}(\vec x; a_{k})=Q^+(a_{k})\Psi^{(1)}_{E_{n,m}}(\vec x; a_{k})
=Q^+(a_{k})Q^+(a_{k-1})\cdots Q^+(a_0)\Psi^{(1) A}_{E_{n,m}}(\vec x; a_0)
\]
(if normalizable) are the wave functions of $ H^{(0)}(\vec x; a_k)$
with energies $E_{n,m}=-\alpha^2(s_n^2+s_m^2)$.
The symmetries of wave functions
alternate and depend on the length of chain. This is true but up to zero modes of
operators~$Q^+$.

It is necessary to keep under the control normalizability of $\Psi$ and zero modes of~$Q^+$.
This control is performed algebraically by means of identity, which must be fulf\/illed up to a function of~$H$
\[
R^{(1)}(a_{k})=R^{(0)}(a_{k-1}).
\]
Actually, the following equation holds:
\[
Q^-(a_{k})Q^+(a_{k})=Q^+(a_{k-1})Q^-(a_{k-1})+\alpha^2(2k+1)
\big[2H^{(0)}(\vec x; a_{k-1})+\alpha^2(2k^2+2k+1)\big].
\]
These relations allowed to evaluate the norms of wave functions.
The result is the following. The spectra of Hamiltonians
$H^{(0)}(\vec x; a_k)$ are not degenerate.
They consist of the bound states with energies $E_{n,m}$,
with indices $|n-m|>k+2$,
and their wave functions $\Psi^{(0)}_{E_{n,m}}(\vec x; a_k)$ were given
analytically above.

\subsection*{Acknowledgements}

I am very grateful to A.A.~Andrianov, F.~Cannata and D.N.~Nishnianidze for fruitful
colla\-boration and many useful discussions. The work was partially supported by the RFFI grant \mbox{09-01-00145-a}.

\pdfbookmark[1]{References}{ref}
\LastPageEnding

\end{document}